%% file: 1550_RPM_MNRAS.tex
\documentclass[useAMS,usenatbib]{mn2e}
\usepackage{graphicx} 
\usepackage{times}
\usepackage{epsfig}
\usepackage{rotating}
\usepackage{sidecap}
\usepackage{longtable}
\usepackage{lscape}
\usepackage{color}
\include{def_bibtex}

\include{definitions}

\title[The spin of the BH in XTE J1550-564 through the RPM]{Black hole spin measurements through the relativistic precession model: XTE J1550-564}
\author[S. Motta et al.]{S.E. Motta$^{1}$, T. Mu\~noz-Darias$^2$, A. Sanna$^3$, R. Fender$^2$, T. Belloni$^4$, L. Stella$^5$ \\%, T.M. Belloni$^2$, L.Stella$^3$, R. Fender$^4$\ \
$^{1}$ESAC, European Space Astronomy Centre, Villanueva de la Ca\~nada, E-28692 Madrid, Spain\\
$^{2}$University of Oxford, Department of Physics, Astrophysics, Denys Wilkinson Building, Keble Road, OX1 3RH, Oxford, United Kingdom
$^{3}$Dipartimento di Fisica, Universit\'a degli Studi di Cagliari, SP Monserrato-Sestu km 0.7, 09042 Monserrato, Italy \\
$^{4}$INAF-Osservatorio Astronomico di Brera, Via E. Bianchi 46, I-23807 Merate, Italy\\
$^{5}$INAF-Osservatorio Astronomico di Roma, Via Frascati 33, I-00040, Monteporzio Catone, Italy\\
}

\begin{document}
\maketitle
\begin{abstract}
We present a systematic analysis of the complete set of observations of the black hole (BH) binary XTE J1550-564 obtained by the \textit{Rossi X-ray Timing Explorer}. We study the fast time variability properties of the source and determine the spin of the black hole through the relativistic precession model. Similarly to what is observed in the BH binary GRO J1655-40, the frequencies of the QPOs and broad band noise components match the general relativistic frequencies of particle motion close to the compact object predicted by the relativistic precession model. The combination of two simultaneously observed quasi-periodic oscillation (QPO) frequencies together with the dynamical BH mass from optical/infrared observations yields a spin equal to a = 0.34 +/- 0.01, consistent with previous determinations from X-ray spectroscopy.  Based on the derived BH parameters, the low frequency QPO emission radii vary from $\sim 30$ gravitational radii ($R_\mathrm{g}$) to the innermost stable orbit for this spin ($\sim 5$ $R_\mathrm{g}$), where they sharply disappear as observed for the case of GRO J1655-40. 

\end{abstract}

\section{Introduction}

Quasi periodic oscillations (QPOs) are commonly observed in the light curves from accreting compact objects and are thought to arise from the innermost regions of the accretion flow. In a power density spectrum (PDS) they take the form of relatively narrow peaks yielding accurate centroid frequencies that can be associated with dynamical motion and/or accretion related timescales in the presence of a gravitational field. 
Even though QPOs have been known for several decades and several authors proposed  models to describe their origin [some of them involving the predictions of the Theory of General Relativity (GR), e.g. \citealt{Esin1997}, \citealt{Titarchuk1999}, \citealt{Tagger1999}, \citealt{Stella1998}, \citealt{Lamb2001},  \citealt{Abramowicz2001}, \citealt{Ingram2011} and references therein], there is no consensus about their nature.

In black-hole systems, low-frequency ($\sim$ 0.1--30 Hz) QPOs (LFQPOs) of different kinds (dubbed type-A, -B and -C QPOs, see e.g. \citealt{Casella2005} and \citealt{Motta2011}) have been detected. They usually show varying centroid frequencies and are often associated with broader peaked noise components (at $\sim$ 1--100 Hz, see e.g. \citealt{Belloni2011} for a review). 
QPOs with even higher frequencies (up to 450 Hz, high-frequency QPOs, HFQPOs) have also been detected (see, e.g. \citealt{Strohmayer2001}), but only in a much smaller number of observations and sources. (\citealt{Belloni2012}). They are usually divided into two classes, known as lower and upper HFQPOs.

Bound orbits of matter in a gravitational field are characterized  by three different frequencies: the orbital  frequency and the vertical and radial epicyclic frequencies. GR predicts that the motion of matter within a few tens of gravitational radii  ($R_g = GM/c^2$) from black holes (BHs) carries the signatures of strong-field gravity effects. QPOs may provide the most promising prospects to measure such characteristic frequencies in the electromagnetic radiation emitted by the matter accreting onto a compact object. The relativistic precession model (RPM, \citealt{Stella1998}, \citealt{Stella1999}, \citealt{Stella1999a}, \citealt{Motta2013}) associates three different QPOs to a combination of the fundamental frequencies of particle motion. Type-C QPOs are associated with the nodal precession frequency (or Lense-Thirring frequency), while the periastron precession frequency and the orbital frequency correspond to the lower and upper HFQPOs, respectively.
\cite{Motta2013} applied the RPM to data of the BH binary GRO J1655-40 and showed that it provides a natural interpretation for the QPOs and broad noise components observed in the PDS of the source. 
\cite{Motta2013} also showed that the RPM constitutes an effective tool to precisely determine both the mass and the spin of a black hole at the same time. Here, we use the same method to measure the spin of the BH binary XTE J1550-564, for which the dynamical mass has been recently measured through optical and infrared (OIR)  observations. (\citealt{Orosz2011}).

%%_____________________BEGIN________TABLE 2____________________________%%
\begin{table} 
\centering 
\caption{Parameters of the simultaneous QPOs detected in the PDS from Obs. 30191-01-02-00 and used in this study. Note that a type B QPO is also present in the PDS.}\label{tab:QPOs} 
\begin{tabular}{c c c c} 
\hline													
QPO type			&	Frequency				&		Width		&		Normalization						\\
				&	$[Hz]$					&		$[Hz]$		&		$Leahy$								\\
\hline													
\hline													
type-B			&	4.92	$^{+0.03}_{-0.02}$	&		0.33	$^{+0.09}_{-0.09}$	&		0.57	$^{+0.03}_{-0.02}$	\\
type-C			&	13.08	$\pm$0.08		&		5.20	$^{+0.27}_{-0.26}$	&		8.08	$^{+0.03}_{-0.02}$	\\
High-frequency	&	183	$\pm$5				&		84	$^{+22}_{-17}$		&		6.1	$^{+1.0}_{-0.9}$		\\
\hline													
\end{tabular}
\end{table} 
%%_____________________END__________TABLE_2____________________________%%
%
%
\section{Observations and data analysis}\label{sec:observations} 

We examined a total of 361 Rossi X-Ray Timing Explorer (RXTE)/Proportional Counter Array (PCA) archival observations\footnote{http://heasarc.gsfc.nasa .gov/docs/xte/archive.html} of XTE J1550-564 obtained between 1998 September 7 (MJD 51063) and 2004 June 6 (MJD 53162). 
The PCA data modes employed for most of these observations include a single-bit mode covering the absolute channel range 0--35 and a high-time-resolution event mode recording events above PCA absolute channel 36.
For each observation we computed PDS using custom software under \textsc{IDL} (GHATS\footnote{http://www.brera.inaf.it/utenti/belloni/GHATS\_Package/Home.html}) following the same method and extractions parameters described in \cite{Motta2013}. We produced PDS in the 1.51-27.40 keV, 1.51-9.52 keV and 9.52-27.40 keV energy bands (corresponding to absolute channels 0--102, 0--35 and 36--102 at the beginning of the RXTE mission), to which we refer as total, soft and hard energy band. We used 64 and 128 seconds-long intervals of the event files and a Nyquist frequency of 2048 Hz to produce the PDS. 
Then, we averaged the individual PDS (obtaining a single PDS for each observation) and subtracted the contribution due to Poissonian noise (see \citealt{Zhang1995}).
The PDS were normalized according to \cite{Leahy1983} and converted to square fractional rms \citep{Belloni1990}. We measured the integrated fractional rms (defined as the rms integrated over a certain frequency band) in the 2-27 keV integrating the PDS over the 0.1--64 Hz frequency band and taking the square root of the power obtained (\citealt{Munoz-Darias2011}). 

PDS fitting was carried out with the {\sc xspec} package by using a one-to-one energy-frequency conversion and an unit response. Following \cite{Belloni2002}, we fitted the noise components with a number of broad Lorentzian shapes and both low frequency and high-frequency QPOs with a narrow Lorentzian. In the case of LFQPOs we added a variable number of Lorentzians depending on the presence of harmonic peaks. 
We measured the frequencies of the narrow features using the PDS produced from 64s-long intervals (because of their higher signal-to-noise), whereas the PDS from 128s long intervals were used for the broader components since they provide a lower  minimum observable frequency.

Based on the results of the fitting, we excluded from the analysis non-significantly detected features (significances\footnote{The single trial signiﬁcances of QPOs are given as the ratio of the integral of the power of the Lorentzian used to fit the QPO divided by the negative 1-sigma error on the integral of the power.} $\leq$ 3$\sigma$, \citealt{Boutelier2010}). The best fit parameters from all the observations that were considered are listed in Tab. \ref{tab:sample}.
\begin{figure}
\centering
\includegraphics[width=7.5cm]{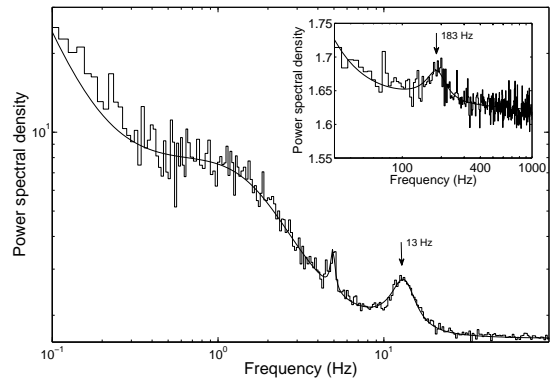}
\caption{PDS obtained from Obs. 30191-01-02-00. The two simultaneously detected QPOs are marked (see large panel for the type-C QPO and the inset for the HFQPO). A narrow type-B QPO is also visible at $\sim$5 Hz }\label{fig:PDS}
\end{figure}
%----------------------------------------------------------------------------------
%
%
\subsection{QPO classification and sample selection}\label{sec:QPO_cla}

Type-C QPOs represent the BH equivalent of the so-called Horizontal Branch Oscillations (HBOs) observed in accreting neutron stars (\citealt{Casella2005}, but see \citealt{Altamirano2012a} for the case of HBOs in Terzan 5 X-2). Therefore Type C QPO are suited to be used in the framework of the RPM (\citealt{Motta2013}). Following \cite{Motta2012}, we selected 49 observations showing type-C QPOs covering the 0.1-18 Hz frequency range (see Tab. \ref{tab:sample}).
Among this sample, we searched for narrow features at high frequencies ($> 100$ Hz). We found only one observation that showed a HFQPO simultaneously to a type-C QPO (Obs Id 30191-01-02-00), both observable in the PDS of the total (see Fig. \ref{fig:PDS}), soft and hard energy band. 
Non-simultaneous HFQPOs at $\sim$180 Hz and $\sim$280 Hz have been observed in XTE J1550-564 by \cite{Remillard1999a} and \cite{Miller2001}, later also reported by \cite{Belloni2012}, who performed a systematic analysis of high-frequency QPOs in a sample of BH binaries.  \cite{Miller2001} reported the simultaneous detection of two HFQPOs (at $\sim$270 Hz and $\sim$170 Hz, respectevely), but unfortunately the latter was statistically not significant (\citealt{Belloni2012}).
\citet{Mendez2013} systematically studied the phase lags of all the HFQPOs detected in XTEJ1550-564 and found that they correspond to a single lower HFQPO that changes frequency over time and that is observed between $\sim$180 and $\sim$280 Hz. Based on this work, we treated the peak at $\sim$183 as a lower HFQPO.
Observation 30191-01-02-00 also shows a peak at $\sim$5 Hz that we classified as a type-B QPO. The two QPOs were classified based on the relation between rms and QPO centroid frequency, following the steps detailed in \citealt{Motta2012}. This is the second case in which simultaneous type-C and type-B QPOs are reported (see \citealt{Motta2012} for GRO J1655-40). The parameters of the QPOs in Obs. 30191-01-02-00 are reported in Tab. \ref{tab:QPOs}.
\section{Results}
\subsection{From the Reltivistic precession model to the spin measurement}\label{sec:RPM_solve}

The functional form of the RPM equations depends solely on the BH mass and spin and the radius at which the QPOs are produced. The system of equations can be solved whenever the three QPOs are observed simultaneously. This condition is not satisfied for XTE J1550-564, where only a type-C QPO and a HFQPO are simultaneously detected. 
However, if we use the BH mass as recently determined from optical spectro-photometric observations (M = 9.10 $\pm$ 0.61 M$_{\odot}$, \citealt{Orosz2011}), the emission radius and BH spin remain the only unknowns in the application of the RPM to the QPOs of XTE J1550-564 and the model equations can thus be solved.
We proceeded as follows:

\begin{enumerate}

\item  For the nodal and the periastron precession frequencies ($\nu_{per}$ and $\nu_{nod}$ in eq. 7, \citealt{Motta2013}) we used the Newton-Raphson method to calculate the set of radii at which the observed frequencies (type-C QPO and lower HFQPO, respectively) are produced for every possible combination of mass and spin in a given range. We considered masses between 3 and $50 \sol$ with a resolution of $0.01 \sol$ and spins between 0 and 1  with a resolution of 0.001. This results in two independent sets of mass-spin-radius solutions. 

\item We inferred the spin that for a mass fixed at the OIR value, solves simultaneously the two equations equations of step (i). 

\item Through the Monte-Carlo method we simulated 10$^5$ sets of two frequencies and BH mass. The distributions are Gaussian, centred at the mean value and with a width equal to the error on the centroid frequencies and the mass, respectively.

\item We solved the RPM system following step 1 and 2. As a result we obtained two distributions of spins and emission radii consistent with being Gaussian-distributed.

\item Fitting the distribution of spin and radius from step (iv), we obtained the following measurements: spin a/M =  0.341$\pm$0.007, radius = (5.47$\pm$0.12) $R_\mathrm{g}$. The best fit parameters are shown in Tab. \ref{tab:distribution_par}.

\item From the spin and the BH mass distributions we obtained a distribution of values for $R_\mathrm{ISCO}$ (see \citealt{Motta2013}) and its corresponding nodal frequency. We find $R_\mathrm{ISCO} = (4.83 \pm 0.02$) $R_\mathrm{g}$ for XTE J1550-564.
\end{enumerate}
%
%%_____________________BEGIN________DISTPAR____________________________%%
%
%
\begin{table}
\centering
\caption{Black hole parameters and associated quantities as measured from the RPM. For details, see the text, Sec. \ref{sec:RPM_solve}. *The BH mass is taken  from the literature and it comes from Optical-Infrared dinamical studies (\citealt{Orosz2011}).\bigskip}\label{tab:distribution_par}						
\begin{tabular}{|c c c|}	

\hline                                                                                                                                                                                                                                            

 	&	Mean value	&	Standard deviation	\\	
\hline                                                                                                                                                                                                                                                  						
\hline                                                                                                                                                                                                                                                  						
Mass* (Solar masses)	&	9.1	&	0.6	\\	
Spin	&	0.341	&	0.007	\\	
Radius (Gravitational radii)	&	5.47	&	0.12	\\	
\hline						
$R_\mathrm{ISCO}$ (Gravitational radii)	&	4.83&	0.02	\\	
$\nu_{nod}$ at $R_\mathrm{ISCO}$ (Hz)	&	18	&	1	\\	
\hline                                                                                                                                                                                                                                                	
\end{tabular}						
\end{table} 						
% 
%%_____________________END__________DISTPAR____________________________%%
%
%----------------------------------------------------------------------------------
\begin{figure}
\centering
%\begin{rotate}{90}
\includegraphics[width=7.5cm]{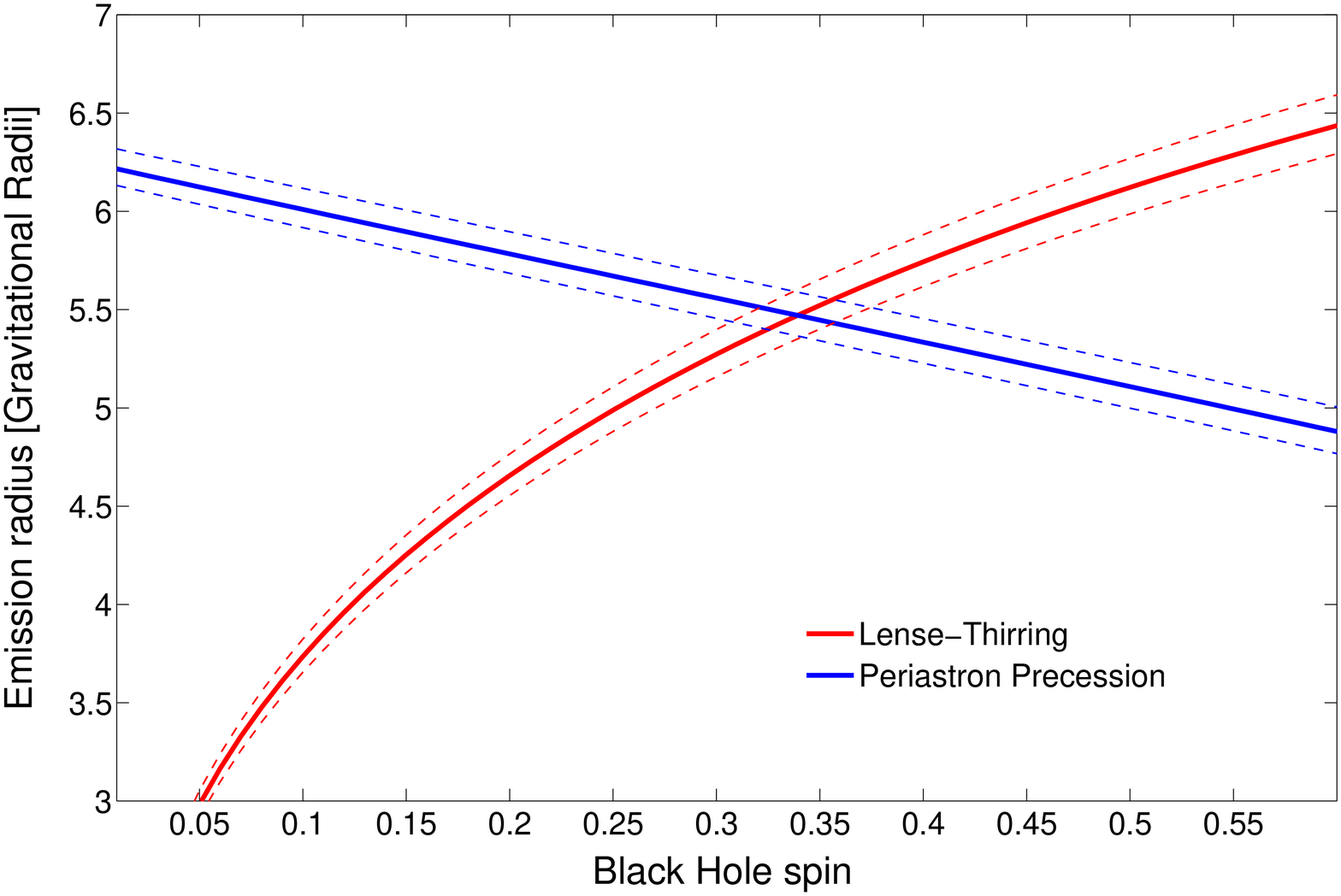}
\caption{Spin as a function of the emission radius as predicted by the two relevant  equations of the relativistic precession model (blue line for nodal precession frequency; red line for periastron precession frequency) for the black hole binary XTE J1550-564. The derived black hole spin is a = 0.34$\pm$0.01. The solid line marks the measured values and the dashed lines mark the 1-sigma confidence level. For the latter, only the error in the BH mass (which is dominant) is considered. }\label{fig:intersezione}
%\end{rotate}
\end{figure}
%----------------------------------------------------------------------------------
%
%
%
\subsection{The PBK correlation and the distribution of the type-C QPOs}\label{sec:PBK}

In the PDS of BHs and NSs X-ray binaries, the frequency of two power-spectral components follow a tight correlation over a very large frequency range (\citealt{Psaltis1999}). This correlation involves either two QPOs (a low-frequency QPO and either the lower or the upper HFQPO) or a LFQPO and a broad noise component. Following \cite{Stella1999} and \cite{Motta2013}, we inspected the observations of our sample to identify the power-spectral component following the PBK correlation (L$_{lf}$, L$_{l}$ and L$_{u}$ according to \citealt{Belloni2002}). 

\begin{figure*}
\centering%\begin{rotate}{90}
\includegraphics[width=16.6cm]{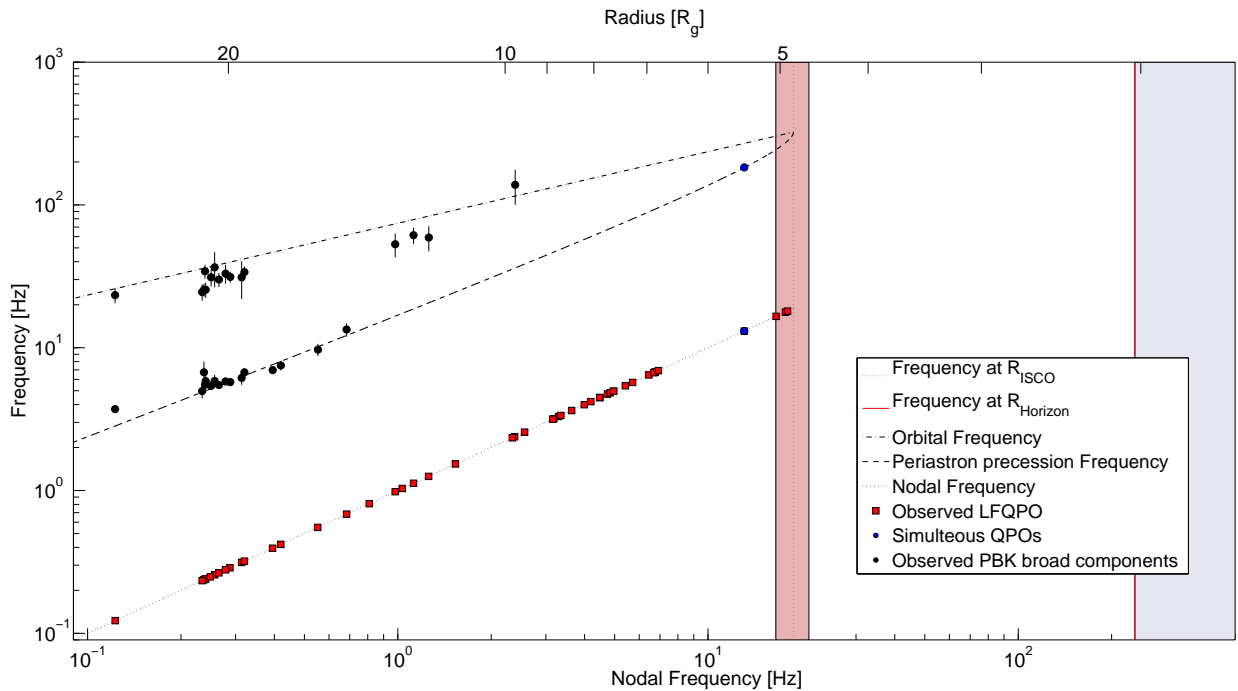}
\caption{Nodal frequency (dotted line), periastron precession frequency (dashed line) and  orbital frequency (dot-dashed line) as a function of the nodal precession frequency around a Kerr black hole as predicted by the relativistic precession model for QPOs. The lines are drawn for mass M = 9.10 M$\odot$ and spin a/M = 0.34. The corresponding radii are given in the top X-axis. The blue points, the black circles and the red squares represent the simultaneous lower HFQPO and type-C QPO, the characteristic frequencies of the broad components following the PBK correlation and the type-C QPOs frequencies plotted against themself, respectively. The vertical dotted red line marks  the nodal frequency produced at the innermost stable circular orbit and the red vertical band indicates its corresponding 3-sigma uncertainty.}\label{fig:nu_vs_nu}
%\end{rotate}
\end{figure*}

We considered the characteristic frequency $\nu_{max}$\footnote{$\nu_{max}$ is defined as ($\nu_{max}^2 = \nu^2 + (\Delta/2)^2)$, where $\Delta$ is the width of the Lorentzian component describing a given power-spectral feature (\citealt{Belloni2002}).} of the components L$_{l}$ and L$_{u}$ and the centroid frequency L$_{lf}$ of the type-C QPO (\citealt{Motta2013}). We plotted the characteristic frequencies L$_{l}$ and L$_{u}$ as a function of the LFQPO centroid frequency. We also plotted the frequencies predicted by the RPM using the  OIR mass and the spin and emission radius obtained in Sec. \ref{sec:RPM_solve}. The result is shown in Fig. \ref{fig:nu_vs_nu}. 

\begin{itemize}

\item The characteristic frequencies of the L$_{l}$ and L$_{u}$ components follow fairly closely the frequencies predicted by the RPM. The L$_{l}$ frequencies approximately match the periastron precession frequency, while the L$_{u}$ frequencies follow the orbital  frequency (see Fig. \ref{fig:nu_vs_nu}). Most L$_{l}$ and L$_{u}$ components centroid frequencies are consistent within 3  $\sigma$ with the frequencies predicted by the RPM. Only two points are  marginally consistent. 

\item Type C QPOs are observed to vary over a broad frequency range. All of them are consistent with being produced at radii larger than $R_\mathrm{ISCO}$. The highest frequency Type-C QPO is centred at 18.04 Hz, which would correspond to a radius equal to 4.9 gravitational radii, only $\sim$1.6\% larger than $R_\mathrm{ISCO}$. 
\end{itemize}
\subsection{On the width of the QPOs}\label{sec:width}

The QPOs observed in BH X-ray binaries are typically narrow, with very small fractional widths (\citealt{vanderKlis1997}), which provide an indication on the geometrical structure of the region where they are produced. As noted by \cite{Motta2013}, the simplest assumption we can make is that the QPOs are produced in a narrow annulus in the accretion flow. To obtain a rough estimate of the radial size of this annulus, we apply a variable jitter $dr$ to the emission radius at which the QPOs are produced. Following \cite{Motta2013} we started with a very small jitter and we increased $dr$ until we obtained a distribution of RPM frequencies with FWHM consistent with the FWHM of the observed QPOs. We see that a 
jitter between 5.3\% and 5.7\% of the emission radius is able to reproduce the widths of the two simultaneous QPOs observed in XTE J1550-564. 
%
%
%----------------------------------------------------------------------------------
%----------------------------------------------------------------------------------
%
\section{Discussion}\label{sec:discussion}

We have examined 361 RXTE archival observations of the BH binary XTE J1550-564 from its 1998 and 2000 outburst and in one case we simultaneously detected a type-C QPO and a HFQPO. Following \cite{Motta2013}, we applied the RPM to XTE J1550-564 by combining the centroid frequencies of the QPOs together with the BH mass inferred from OIR observations. We obtained a spin of a/M = 0.34 $\pm$ 0.01 and an emission radius for the QPOs of $(5.5 \pm 0.1) R_\mathrm{g}$. 
Assuming that the two simultaneous QPOs (Type-C and lower HFQPO) arise from an annular region around the emission radius, assuming a jitter of the emission radius we estimate the width of this annulus to be $\pm 5-6\%$ of the emission radius. 
 This is in agreement with the results by \cite{Heil2011}, who showed that the Type-C QPOs in XTE J1550-564 display frequency jitter on short timescales (seconds) and also  reported that the frequency of the Type-C QPO correlates with the flux over these timescales. This quantitatively supports our treatment of the radius jitter since for larger radii (R+dr) we expect lower QPO frequencies and fluxes than for smaller radii (R-dr).

As a check, we also solved the RPM system assuming that the HFQPO at $\sim$183Hz corresponds to the orbital frequency (instead of the periastron precession frequency). We obtained a spin of  a/M = 0.84 and an emission radius of $7 R_\mathrm{g}$. We plotted the frequencies predicted by the RPM using these values and the characteristic frequencies L$_{l}$ and L$_{u}$ (see Sec. \ref{sec:PBK}), but we found that there is no agreement between the observed and the predicted frequencies. Therefore, we can exclude that the HFQPO at $\sim$183 Hz corresponds to the orbital frequency.

\cite{Steiner2011} reported a measurement of the spin of the BH in XTE J1550-564  by modelling both the thermal continuum spectrum of the accretion disc and by fitting the Fe-K$\alpha$ line. They obtained $-0.11 < a/M < 0.71$ (90 per cent confidence) for the former method, whereas the latter results in a/M = 0.55$^{+0.15}_{-0.22}$. Their combined result is a/M = 0.49$^{+0.13}_{-0.20}$, consistent with the value reported in this work. %However, we note that our measurement better fits with the spin measurement coming from the spectral continuum modeling that, despite being less precise than the iron line method, provided a most likely spin of a/M = 0.34. %\cite{Svoboda2009} found that BH spin measurements from  broadened Iron lines depend on the distribution of the disc emission  assumed during the fitting. In particular, an inproper use of the directionality proﬁle can affect the  parameters inferred for the relativistic broad line model. Especially for the often used case of limb darkening, the radial steepness can interfere with the line parameters of the best-ﬁt model by enhancing the red wing of the line and resulting in an over estimation of the BH spin.  

In \cite{Motta2013} we found that the spin value from the RPM to the data of GRO J1655-40 disagreed with the value obtained from both the thermal continuum modelling and the Fe-K$\alpha$ line fitting. However, we note that despite the fact that the orbital inclination of GRO J1655-40 and XTE J1550-564 are comparable (both have been estimated between 70 and 75 degrees, \citealt{Orosz2002}, \citealt{Greene2001}), \cite{Greene2001} found that in GRO J1655-40 the jet axis -  which is presumably aligned with the BH spin axis - and the orbital axis are misaligned by $> 15^{\circ}$. On the contrary, \cite{Steiner2012b} determined the inclination angle of the jet axis in XTE J1550-564 and found that it is probably aligned with the BH spin axis (they infer an upper limit to the BH spin-jet axis misalignment of 12$^{\circ}$). Therefore, while the assumption that the spin inclination to the line of sight is the same as the orbital inclination seems correct within small errors in the case of XTE J1550-564, this might not be true for GRO J1655-40. 
%{\bf A simple reasoning supports this conclusion. In the case of GRO J1655-40, the continuum method provides a measurement of the ISCO radius R$^{con}_\mathrm{ISCO} = 3.61$, which is proportional to $[cos(i_{out})]^{-0.5}$ (see \citealt{Kubota1998}), where $i_{out}$ is the orbital inclination (the angle between the line of sight and the disk axis). Through the RPM we obtain R$^{RPM}_\mathrm{ISCO} = 5$, which is instead proportional to $[cos(i_{in})]^{-0.5}$, where $i_{in}$ is the inclination between the line of sight and the BH spin axis. This implies that $(Rrpm/Rcon)^2 = cos(i_{out}) / cos(i_{in})$, which - assuming $i_{out} = 70$ degrees - gives $i_{in} ~ 83$ degrees. This would explain why the QPOs are so powerful in GRO 1655-40 (see \citealt{Schnittman2006}, Fig. 4 ).}
In any case, it is clear that more timing measurements are needed to test whether the RPM-based method and the two spectroscopic methods are generally consistent [note that in some cases the spectroscopic methods already disagree \citep{Reynolds2013}].  However, since currently available data do not provide the adequate signal to noise to further probe this apparent inconsistency, it is clear that only new generation satellites such as LOFT (\citealt{Feroci2011}) will be able to tell us more about HF QPOs and to solve the issue.

The spin value obtained from the RPM, together with the mass estimated from the OIR allow us to predict the  expected behaviour of the frequencies for each QPO type in XTE J1550-564. All frequencies reach their highest theoretically allowed values at the innermost stable circular orbit, where the relativistic effective potential has an inflection point and the orbital and periastron precession frequencies coincide (at $\sim$320 Hz). Based on the mass and spin values that we measure, the ISCO is expected at a radius of 4.8 $R_\mathrm{g}$, corresponding to a nodal precession frequency of 18.8 Hz. 

Once the RPM is solved, the identification of any of the three frequencies relevant to the RPM at other times allows us to measure the corresponding emission radius. In many of the RXTE observations, XTE J1550-564 displayed low-frequency QPOs (Fig. \ref{fig:nu_vs_nu}) whose frequency varied over  a wide range, covering about two decades, from $\sim$0.1 Hz to $\sim$18 Hz \citep{Motta2012}. The lowest  frequency type-C QPO is produced at $\sim$30 gravitational radii from the BH, while the highest arises from a radius consistent with the ISCO radius. A similar result was found for the case of GRO J1655-40.

As already noted by \cite{Stella1999a}, the RPM also provides a natural interpretation for the PBK correlation (\citealt{Psaltis1999}), which involves either two QPOs (a low-frequency QPO and either the lower or the upper HFQPO) or a low-frequency QPO and a broad noise component. Following \cite{Motta2013}, the periastron precession and orbital frequencies as a function of the nodal frequency match the correlation between the type-C QPOs and the broad noise components frequencies observed in GRO J1655--40 (see Fig. \ref{fig:nu_vs_nu}). 
Differently from what we saw in GRO J1655-40, in XTE J1550-564 we simultaneously observe the L$_l$ and the L$_u$ components (associated to the periastron precession and to the orbital motion, respectively, in the RPM framework). 
We note that, in general, the centroid frequencies of the L$_l$ and the L$_u$ components display some scatter around the RPM predicted frequencies. Furthermore, the L$_u$ characteristic frequencies tend to be shifted to lower frequencies than the predicted frequencies. We identify three different reasons for this:
\begin{itemize}

\item Deviations of the behaviour of the matter in the accretion flow from the test-particle description are expected especially at large radii (i.e. when the type-C QPO shows a centroid frequency below $\sim$1), when the region in which both QPOs and broad frequencies originate is probably radially larger than at smaller radii.

\item Differently from QPOs (which are relatively narrow),  the precision with which the characteristic frequencies of the broad PBK components are measured is affected by the model used to fit them. This introduces additional uncertainties that could partly explain the scatter/shift of the points and their relative shift. %Therefore, these uncertainties should be better/more realistically quantified. 
However, a detailed treatment of these uncertainties is beyond the scope of this work and it will be discussed in a forthcoming paper (Motta et al. in prep).

\item The point above is particularly true for the L$_u$ component, which is the hardest to constrain. In general, disentangling the L$_l$ and the L$_u$ components is problematic and constraining well the parameters is often prevented by the low S/N at high frequencies (above $\sim$10Hz, where these broad componets are more powerful).

\end{itemize}
%
%Nevertheless, the agreement between the observed and predicted frequencies is such that, in principle, we could solve the RPM system as described in Motta et al. (2013). A full Monte-Carlo-based solution using these frequencies is beyond the scope of this work and it will be discussed in a forthcoming paper (Motta et al. in prep). To illustrate this we have solved the RPM system only using the central value of these frequencies (and therefore not propagating their errors). We obtain masses between 11 and 25 M$_{\odot}$ and spins between 0.21 and 0.32.
%
%\subsection{Jets and spins}
%Radio jets are ubiquitously observed in BH binaries (\citealt{Fender2004}), and have been proposed to be powered by BH spin (\citealt{Narayan2012}). Since the spins of XTE J1550-564 and GRO J1655-40  - measured through the RPM - are  similar (a = 0.34 and a = 0.28, respectively), we would expect similar jet powers for them unless the spin-jet dependence is very strong. However, we note that irrespectively of using either the jet power (\citealt{Fender2010}) or the radio peak luminosity (\citealt{Narayan2012}) the values obtained for XTE J1550-564 and GRO J1655-40 differ by about an order of magnitude (larger for the latter). Hence, they are inconsistent with spin powering the jet if our RPM spin measurements are correct. We note that XTE J1550-564 and GRO J1655-40 are 2 out of the 4 systems used in the spin-jet correlation presented in \citet{Narayan2012}.
%
%Observing the same properties already observed in GRO J1655-40 also in XTE J1550-564 strenghten the validity of the RPM. 
%
\section{Summary and conclusions}								

XTE J1550-564 is the second black hole binary whose timing properties can be interpreted in terms of motion of matter in the close vicinity of a rotating BH. Here, we report a spin measurement for the BH in this source obtained by applying the relativistic precession model and based on the method presented in \cite{Motta2013}. We combine the results from X-ray timing (i.e. the frequency of two different simultaneously observed oscillations) with the dynamical BH mass obtained from OIR observations, yielding spin and emission radius measurements with formal errors as small as $\sim$2\% (a/M = 0.34 $\pm$ 0.01), the former being consistent with previous results obtained through X-ray spectroscopy. 

The agreement between the predictions of the RPM and the actual timing properties of XTE J1550-564 strengthen the conclusions reported by \cite{Motta2013} and further supports the hypothesis that both the QPOs and broad noise components in accreting BH binaries could, on average, be explained in terms of motion of matter in the close vicinity (a few tens $R_\mathrm{g}$ down to a few $R_\mathrm{g}$) of a BH. 

\bigskip

\small
\noindent The authors acknowledge the anonymous referee for his/her helpful comments and suggestions that contributed to improve this work. SEM acknowledges the ESA research fellowship program. SEM also acknowledges Jiri Svoboda for useful discussions. TMD acknowledges funding via an EU Marie Curie Intra-European Fellowship under contract no. 2011-301355. SM and TMB acknowledge support from INAF PRIN 2012-6
This research has made use of data obtained from the High Energy Astrophysics Science Archive Research Center (HEASARC), provided by NASA's Goddard Space Flight Center. 
\normalsize
\bibliographystyle{mn2e.bst}
\bibliography{biblio.bib}

\onecolumn																													
\begin{center} 																													
%\begin{landscape}																													
\begin{longtable}{|c c c c |c|c c|} 																													
\caption{Observations included in our sample. We report the Observation Id, the time at which the observation was taken, the integrated fractional rms (measured in the 2-27 keV energy band and in the 0.0-64.0 Hz frequency range), the hardness ration calculated as described in Sec. \ref{sec:observations}, the centroid frequency of the Type-C QPO observed, the characteristic frequency ($\nu_{Max}$) of the broad high-frequency component - when detected. For details on the model used to fit the PDS, see Sec. \ref{sec:observations}}\label{tab:sample} \\ 																													
\endfirsthead																													
																													
\multicolumn{7}{c}%																													
{{\tablename\ \thetable{} -- continued from previous page}} \\ \hline																													
\multicolumn{4}{|c|}{ } & \multicolumn{1}{|c|}{Type-C QPOs} & \multicolumn{1}{c}{broad component L$_{l}$} & \multicolumn{1}{c|}{broad component L$_{u}$} \\																													
\hline																								\hline
					
Time	&	Obs. ID	&	Hardness ratio			&	rms 			&	$\nu$					&	$\nu_{Max}$					&	$\nu_{Max}$					\\
$[MJD]$	&		&				&	$[\%]$			&	$[Hz]$					&	$[Hz]$					&	$[Hz]$					\\
																													
\hline																								\hline
					
\endhead																													
\hline \multicolumn{7}{c}{{Continued on next page}} \\																													
\endfoot																													
																									\hline			
																													
\endlastfoot																															
\hline																													
\multicolumn{4}{|c|}{ } & \multicolumn{1}{|c|}{Type-C QPOs} & \multicolumn{1}{c}{broad component L$_{l}$} & \multicolumn{1}{c|}{broad component L$_{u}$} \\																													
\hline																								\hline 					
Time	&	Obs. ID	&	Hardness ratio			&	rms 			&	$\nu$					&	$\nu_{Max}$					&	$\nu_{Max}$					\\
$[MJD]$	&		&				&	$[\%]$			&	$[Hz]$					&	$[Hz]$					&	$[Hz]$					\\
																													
\hline																							\hline				
																													
51065.1	&	30188-06-01-00	&	0.948	$\pm$	0.003	&	31.0	$\pm$	0.3	&	0.288	$_{-	0.001	} ^{+	0.001	}$   &	3.7	$_{-	0.1	} ^{+	0.1	}$   &	23	$_{-	3	} ^{+	3	}$  \\
51065.3	&	30188-06-01-01	&	0.937	$\pm$	0.003	&	28.5	$\pm$	0.3	&	0.395	$_{-	0.002	} ^{+	0.002	}$   &	5.7	$_{-	0.3	} ^{+	0.3	}$   &	31	$_{-	3	} ^{+	2	}$  \\
51066.1	&	30188-06-01-02	&	0.892	$\pm$	0.003	&	26.2	$\pm$	0.2	&	0.809	$_{-	0.002	} ^{+	0.002	}$   &	7.0	$_{-	0.3	} ^{+	0.3	}$   &	 -					\\
51066.3	&	30188-06-01-03	&	0.865	$\pm$	0.003	&	26.5	$\pm$	0.2	&	1.034	$_{-	0.003	} ^{+	0.003	}$   &	9.1	$_{-	0.5	} ^{+	0.5	}$   &	 -					\\
51064.0	&	30188-06-03-00	&	0.968	$\pm$	0.003	&	32.1	$\pm$	0.3	&	0.123	$_{-	0.001	} ^{+	0.001	}$   &	 -					&	 -					\\
51067.3	&	30188-06-04-00	&	0.798	$\pm$	0.002	&	26.5	$\pm$	0.2	&	1.534	$_{-	0.004	} ^{+	0.004	}$   &	 -					&	 -					\\
51068.3	&	30188-06-05-00	&	0.701	$\pm$	0.002	&	23.4	$\pm$	0.2	&	2.381	$_{-	0.005	} ^{+	0.005	}$   &	 -					&	138	$_{-	38	} ^{+	47	}$  \\
51069.3	&	30188-06-06-00	&	0.623	$\pm$	0.002	&	20.6	$\pm$	0.1	&	3.297	$_{-	0.006	} ^{+	0.006	}$   &	 -					&	 -					\\
51070.1	&	30188-06-07-00	&	0.632	$\pm$	0.002	&	20.9	$\pm$	0.1	&	3.182	$_{-	0.004	} ^{+	0.004	}$   &	 -					&	 -					\\
51070.3	&	30188-06-08-00	&	0.634	$\pm$	0.002	&	20.8	$\pm$	0.1	&	3.161	$_{-	0.003	} ^{+	0.004	}$   &	 -					&	 -					\\
51071.2	&	30188-06-09-00	&	0.601	$\pm$	0.002	&	20.0	$\pm$	0.1	&	3.634	$_{-	0.005	} ^{+	0.005	}$   &	 -					&	 -					\\
51072.0	&	30188-06-10-00	&	0.680	$\pm$	0.002	&	22.3	$\pm$	0.3	&	2.563	$_{-	0.006	} ^{+	0.006	}$   &	 -					&	 -					\\
51072.3	&	30188-06-11-00	&	0.585	$\pm$	0.002	&	19.3	$\pm$	0.1	&	3.992	$_{-	0.005	} ^{+	0.005	}$   &	 -					&	 -					\\
51074.1	&	30191-01-01-00	&	0.533	$\pm$	0.002	&	16.2	$\pm$	0.1	&	5.713	$_{-	0.010	} ^{+	0.010	}$   &	 -					&	 -					\\
51076.0	&	30191-01-02-00 	&	0.600	$\pm$	0.002	&	2.5	$\pm$	0.0	&	13.087	$_{-	0.080	} ^{+	0.080	}$   &	 -					&	 -					\\
51087.7	&	30191-01-18-01	&	0.621	$\pm$	0.002	&	22.5	$\pm$	0.2	&	3.356	$_{-	0.008	} ^{+	0.008	}$   &	 -					&	 -					\\
51096.6	&	30191-01-27-00	&	0.538	$\pm$	0.002	&	18.5	$\pm$	0.1	&	5.414	$_{-	0.009	} ^{+	0.009	}$   &	 -					&	 -					\\
51095.6	&	30191-01-27-01	&	0.573	$\pm$	0.002	&	20.8	$\pm$	0.2	&	4.481	$_{-	0.009	} ^{+	0.009	}$   &	 -					&	 -					\\
51097.8	&	30191-01-28-00	&	0.586	$\pm$	0.002	&	21.8	$\pm$	0.2	&	4.191	$_{-	0.010	} ^{+	0.010	}$   &	 -					&	 -					\\
51097.6	&	30191-01-28-01	&	0.561	$\pm$	0.002	&	20.3	$\pm$	0.1	&	4.742	$_{-	0.009	} ^{+	0.009	}$   &	 -					&	 -					\\
51098.3	&	30191-01-28-02	&	0.558	$\pm$	0.002	&	20.1	$\pm$	0.1	&	4.959	$_{-	0.009	} ^{+	0.009	}$   &	 -					&	 -					\\
51099.2	&	30191-01-29-00	&	0.559	$\pm$	0.002	&	20.2	$\pm$	0.1	&	4.835	$_{-	0.009	} ^{+	0.009	}$   &	 -					&	 -					\\
51099.6	&	30191-01-29-01	&	0.557	$\pm$	0.002	&	19.9	$\pm$	0.1	&	4.961	$_{-	0.007	} ^{+	0.007	}$   &	 -					&	 -					\\
51100.3	&	30191-01-30-00	&	0.513	$\pm$	0.002	&	15.3	$\pm$	0.1	&	6.449	$_{-	0.009	} ^{+	0.009	}$   &	 -					&	 -					\\
51101.6	&	30191-01-31-00	&	0.504	$\pm$	0.002	&	13.9	$\pm$	0.1	&	6.764	$_{-	0.020	} ^{+	0.020	}$   &	 -					&	 -					\\
51101.9	&	30191-01-31-01	&	0.507	$\pm$	0.002	&	14.3	$\pm$	0.1	&	6.718	$_{-	0.012	} ^{+	0.012	}$   &	 -					&	 -					\\
51644.5	&	50137-02-01-00 	&	0.801	$\pm$	0.003	&	31.5	$\pm$	0.2	&	0.237	$_{-	0.002	} ^{+	0.002	}$   &	6.7	$_{-	1.3	} ^{+	1.1	}$   &	25	$_{-	3	} ^{+	4	}$  \\
51646.3	&	50137-02-02-00 	&	0.795	$\pm$	0.003	&	31.6	$\pm$	0.3	&	0.257	$_{-	0.003	} ^{+	0.003	}$   &	5.8	$_{-	0.6	} ^{+	0.6	}$   &	37	$_{-	10	} ^{+	12	}$  \\
51646.6	&	50137-02-02-01 	&	0.799	$\pm$	0.003	&	31.5	$\pm$	0.3	&	0.265	$_{-	0.002	} ^{+	0.002	}$   &	5.5	$_{-	0.3	} ^{+	0.3	}$   &	30	$_{-	3	} ^{+	3	}$  \\
51648.7	&	50137-02-03-00 	&	0.796	$\pm$	0.003	&	32.4	$\pm$	0.3	&	0.239	$_{-	0.002	} ^{+	0.002	}$   &	5.5	$_{-	0.3	} ^{+	0.3	}$   &	34	$_{-	4	} ^{+	5	}$  \\
51649.8	&	50137-02-03-01G 	&	0.803	$\pm$	0.004	&	32.7	$\pm$	0.4	&	0.240	$_{-	0.002	} ^{+	0.002	}$   &	5.8	$_{-	0.5	} ^{+	0.5	}$   &	25	$_{-	3	} ^{+	3	}$  \\
51650.7	&	50137-02-04-00 	&	0.796	$\pm$	0.003	&	33.0	$\pm$	0.3	&	0.234	$_{-	0.003	} ^{+	0.003	}$   &	5.0	$_{-	0.5	} ^{+	0.6	}$   &	25	$_{-	3	} ^{+	6	}$  \\
51651.4	&	50137-02-04-01 	&	0.792	$\pm$	0.003	&	32.2	$\pm$	0.2	&	0.249	$_{-	0.002	} ^{+	0.002	}$   &	5.4	$_{-	0.4	} ^{+	0.3	}$   &	31	$_{-	4	} ^{+	4	}$  \\
51652.2	&	50137-02-05-00 	&	0.795	$\pm$	0.003	&	31.9	$\pm$	0.3	&	0.279	$_{-	0.003	} ^{+	0.003	}$   &	5.8	$_{-	0.4	} ^{+	0.3	}$   &	33	$_{-	5	} ^{+	4	}$  \\
51653.5	&	50137-02-05-01 	&	0.787	$\pm$	0.003	&	32.4	$\pm$	0.3	&	0.314	$_{-	0.002	} ^{+	0.002	}$   &	6.1	$_{-	0.6	} ^{+	0.7	}$   &	31	$_{-	9	} ^{+	7	}$  \\
51654.7	&	50137-02-06-00 	&	0.784	$\pm$	0.003	&	32.8	$\pm$	0.3	&	0.320	$_{-	0.002	} ^{+	0.002	}$   &	6.7	$_{-	0.4	} ^{+	0.4	}$   &	34	$_{-	3	} ^{+	3	}$  \\
51655.7	&	50137-02-07-00 	&	0.777	$\pm$	0.003	&	31.8	$\pm$	0.5	&	0.420	$_{-	0.003	} ^{+	0.003	}$   &	7.5	$_{-	0.6	} ^{+	0.5	}$   &	 -					\\
51658.6	&	50134-02-01-00 	&	0.712	$\pm$	0.003	&	29.8	$\pm$	0.4	&	1.257	$_{-	0.006	} ^{+	0.006	}$   &	 -					&	59	$_{-	12	} ^{+	13	}$  \\
51676.4	&	50134-01-04-00 	&	0.443	$\pm$	0.002	&	14.4	$\pm$	0.1	&	6.915	$_{-	0.022	} ^{+	0.022	}$   &	 -					&	 -					\\
51678.5	&	50134-01-05-00 	&	0.550	$\pm$	0.002	&	20.6	$\pm$	0.1	&	4.476	$_{-	0.011	} ^{+	0.011	}$   &	 -					&	 -					\\
51682.3	&	50135-01-02-00 	&	0.669	$\pm$	0.003	&	25.4	$\pm$	0.2	&	2.340	$_{-	0.010	} ^{+	0.010	}$   &	 -					&	 -					\\
51683.8	&	50135-01-03-00	&	0.736	$\pm$	0.003	&	26.1	$\pm$	0.1	&	1.124	$_{-	0.007	} ^{+	0.007	}$   &	 -					&	61	$_{-	8	} ^{+	9	}$  \\
51684.8	&	50135-01-04-00 	&	0.750	$\pm$	0.005	&	25.4	$\pm$	0.3	&	0.981	$_{-	0.020	} ^{+	0.020	}$   &	 -					&	53	$_{-	10	} ^{+	10	}$  \\
51686.3	&	50135-01-05-00 	&	0.761	$\pm$	0.004	&	26.0	$\pm$	0.2	&	0.684	$_{-	0.008	} ^{+	0.008	}$   &	13.4	$_{-	1.4	} ^{+	1.6	}$   &	 -					\\
51687.2	&	50135-01-06-00 	&	0.772	$\pm$	0.005	&	26.1	$\pm$	0.3	&	0.552	$_{-	0.008	} ^{+	0.009	}$   &	9.7	$_{-	0.9	} ^{+	1.3	}$   &	 -					\\
51115.3	&	30191-01-36-00 	&	0.361	$\pm$	0.001	&	4.1	$\pm$	0.1	&	16.576	$_{-	0.224	} ^{+	0.201	}$   &	 -					&	 -					\\
51180.8	&	40401-01-02-00 	&	0.304	$\pm$	0.001	&	2.2	$\pm$	0.0	&	17.766	$_{-	0.121	} ^{+	0.118	}$   &	 -					&	 -					\\
51239.1	&	40401-01-48-00 	&	0.291	$\pm$	0.001	&	2.4	$\pm$	0.0	&	18.037	$_{-	0.067	} ^{+	0.069	}$   &	 -					&	 -					\\

\end{longtable}																													
%\end{landscape}																													
\end{center} 																													
																													
\twocolumn

\label{lastpage}
\end{document}

%% file: def_bibtex.tex
\def\apjs{ApJ}

\def\aap{A\&A}
\def\apjl{ApJ}

%% file: definitions.tex
\def\lesssim{\mathrel{\hbox{\rlap{\hbox{\lower4pt\hbox{$\sim$}}}\hbox{$<$}}}}
\let\la=\lesssim
\def\gtrsim{\mathrel{\hbox{\rlap{\hbox{\lower4pt\hbox{$\sim$}}}\hbox{$>$}}}}
\let\ga=\gtrsim

\def\sol{~\mathrm{M}_\odot}